# Transfer doping of Graphene by Species of Extreme Work Function

*Haichang Lu[1], Yuzheng Guo[2], John Robertson[1]\**

[1]Department of Engineering, Cambridge University, Cambridge CB3 0FA, United Kingdom
[2]College of Engineering, Swansea University, Swansea SA1 8EN, United Kingdom

**E-mail:** jr214@cam.ac.uk

**Abstract:** Density functional calculations are used to explain the charge transfer doping mechanism by which species physisorptively bonded to graphene can increase its free hole or electron density, without giving rise to defects, and thus maintain a high carrier mobility. Typical dopants studied are $FeCl_3$, $AuCl_3$, $SbF_5$, $HNO_3$, $MoO_3$, $Cs_2O$ and $O_2$. These systems do not break the π bonding of the basal plane are particularly important as these do not degrade the carrier mobility. In contrast, more reactive radicals like -OH cause a puckering of the basal plane and thereby act as defects.

**Introduction**

Graphene is a two-dimensional material with a unique band structure with bands crossing at the Dirac point.[1] This gives graphene a very high carrier mobility, but the carrier concentration is small, so that its overall electrical conductivity is rather low.[2] Thus graphene must be doped to increase carrier concentration in order to realize some of its main applications such as a transparent electrode in displays or photovoltaic devices.[2-6] Doping is also useful to increase the sensitivity of graphene when used as a molecular sensor.[7-10]

The conventional way to dope a 3-dimensionally bonded semiconductor would be by substitutional doping. This has indeed been carried out for graphene using nitrogen or boron doping.[9-11] Substitutional sites have the advantage that they are fully bonded into the lattice and are thus stable. However, nitrogen can enter the graphene lattice in various configurations, only one of which is the actual doping configuration.[12,13] The other configurations not only do not dope, they also introduce defects[14,15] which cause carrier scattering and so they degrade the carrier mobility. This 'functionalization' is useful in other contexts such as creating catalytic sites in carbon nanotubes.[14] On the other hand, for graphene as an electrode, it is useful to consider interstitial or transfer doping by physisorbed species.[16-23] These can dope the graphene n-type or p-type, without necessarily creating defects. Transfer doping is also useful to increase the conductivity of contacts, as the high resistance of contacts to graphene in devices can limit the device performance.

We consider the cases of very electronegative and electropositive species, (strong Lewis acids and Lewis bases). These cases are often related to previously studied cases of the intercalation of graphite,[24,25] or the transfer doping of organic molecules for example for organic light emitting diodes.[26,27] We also consider radical species with empty states below the graphene Fermi energy. The second factor of interest is whether the dopant species disturbs the π bonding of the basal plane and thus introduces $sp^3$ sites in the otherwise $sp^2$ bonding. This leads to a loss of carrier mobility.

**Methods**

The calculations are carried out using periodic supercell models of the graphene and the dopant species using the CASTEP plane-wave density functional theory (DFT) code, with ultra-soft pseudopotentials and the Perdew-Burke-Ernzerhof (PBE) form of the generalized gradient approximation (GGA) for the electronic exchange-correlation functional. For the open shell magnetic system $FeCl_3$, we use the GGA+U method, with an on-site potential U of 7 eV applied to the Fe 3d states.

The dispersion correction to the GGA treatment of the van der Waals interaction is included using the Tkatchenko-Scheffler (TS) scheme. To overcome the error induced by periodical mirror charge, self-consistent dipole correction is implemented. The plane wave cut-off energy is 380eV, as the cut-off energy of oxygen.

For the graphene plus dopant system, a layer-by-layer stacked supercell is created in each case, with a close degree of lattice matching between the graphene and the dopant. A 30Å vacuum layer is included where a vacuum layer is needed. The size of the supercell is given in Table. 1. A dense $9 \times 9$ k-point mesh is used, as graphene needs a dense k-point due to its small density of states close to the Dirac point. The calculated lattice constant of graphene in PBE is 2.47Å, 0.4% less than the experimental value.[1]

Doping causes a shift in the system's Fermi energy away from the Dirac point of the graphene. This shift is compared to the Fermi energy, ionization potential or electron affinity of the isolated dopant system. These energies are calculated using a periodic supercell of the dopant species plus vacuum gap. The electrostatic potential is calculated for the dopant system layers, averaged along the layers. The potential in the vacuum gap region gives the vacuum potential. The energy of the valence band maximum is then compared to the vacuum energy to give the ionization potential, and with the band gap, the electron affinity.

**Results**

We first consider Lewis acids such as $AuCl_3$. Crystalline $AuCl_3$ consists of stacked layers of $Au_2Cl_6$ molecular units. The $Au_2Cl_6$ molecule consists of two planar edge-connected $AuCl_4$ squares. The supercell consists of alternate graphene and $AuCl_3$ layers along the z axis. Fig. 2(a) shows the 4x4 graphene supercell with the planar $Au_2Cl_6$ units separated from each other in-plane at a similar distance as in pure $AuCl_3$. The position of $Au_2Cl_6$ units on the graphene is allowed to vary to minimize the total energy.

Fig. 2(c) shows the band structure of isolated pure $Au_2Cl_6$ in the hexagonal lattice. The $Au_2Cl_6$ is a semiconductor with a band gap of 1.22 eV. The Au 5d band is filled to $d^{9.6}$. The conduction band consists of the Au s state and Cl p states. Fig. 2(d) shows the band structure of the combined system. As a 4x4 supercell was used, the graphene Dirac point still lies at K, and can be recognized as the crossed bands at 1.02 eV. This shows that the shift of the Fermi energy $E_F$ due to this $AuCl_3$ doping concentration is 1.02 eV.

Fig. 2(b) shows these results in a density of states (DOS) plot. The doping has occurred by a transfer of electrons from its valence band into the $AuCl_3$ conduction band filling its conduction bands lying just below 0 eV in the central panel of Fig. 2(b).

We next consider $FeCl_3$, which is another Lewis acid like $AuCl_3$. It has been extensively used as an intercalant of graphite, as discussed by Li *et al.*[37] Solid $FeCl_3$ forms a layered system of $Fe_2Cl_6$ edge-connected octahedra connected along three directions at 120° to each other. The Cl sites are rotated slightly off the vertical. The hexagonal supercells of graphene and $FeCl_3$ can be made with a 23Å periodicity. On the other hand, we achieve a smaller $\sqrt{7}\times\sqrt{7}$ supercell using a 1x1 periodicity of the $FeCl_3$ sublattice and a $\sqrt{7}\times\sqrt{7}$ periodicity of the graphene, as in Fig. 3(a). $FeCl_3$ is a magnetic semiconductor with a 0.7 eV band gap. A vertical stacking of one $FeCl_3$ layer and one graphene layer along Oz is ferromagnetic. A stacking of two $FeCl_3$ layers and two graphene layers along Oz as here allows the $FeCl_3$ to be anti-ferromagnetically (AF) ordered, which simplifies the band structure plots (the spin-up and spin-down bands are degenerate). Fig. 3(c) shows the AF bands of isolated $FeCl_3$ calculated for a value of U= 7 eV, with the 0.7 eV band gap. The Fe 3d occupancy is $d^{5.6}$.

Fig. 3(d) shows the band structure of the combined system. The graphene Dirac point can be recognized at the K point 1.0 eV above the Fermi energy. Fig. 3(b) shows the density of states for the combined system and for the isolated $FeCl_3$. Doping has occurred by transfer of electrons from the upper graphene valence band to the $FeCl_3$ conduction states at -0.1 eV in Fig. 3(b).

We then consider the strongest Lewis acid $SbF_5$. Condensed $SbF_5$ can be considered to form a network of corner-sharing octahedral with the F sites vertically above each other. The $SbF_5$ units form chains which are conveniently lattice-matched to graphene, when a supercell of 1x1 $SbF_5$ and $\sqrt{3}\times\sqrt{3}$ is used, as in Fig. 4(a).

Fig. 4(c) shows the band structure of isolated $SbF_5$ in the unit cell of Fig. 4(a). It is a semiconductor with a GGA band gap of 3.06 eV, and a direct gap at $\Gamma$. This system contains only s,p electrons and Sb is in its +5 valence state. The top of the valence band consists of F $2p\pi$ states the conduction band minimum consists of empty Sb 5s states.

Fig. 4(d) shows the band structure of the combined system Due to the orientation of the graphene and $SbF_5$ sub lattices, the Dirac point folds over to appear at $\Gamma$, at about 1.0 eV above the combined Fermi energy. Fig. 4(b) shows the density of states of the combined systems, and the isolated graphene. Doping has occurred by transfer of electrons from the graphene valence band into $SbF_5$ conduction band. This has caused a 3.00 eV shift of the $SbF_5$ bands, but only a 1.05 eV downward shift of the $E_F$ of graphene.

Table. 3 gives the calculated electron affinity, band gap and ionization potential of these compounds. As ideal isolated semiconductors, their Fermi energies would appear near midgap. In practice, the anion vacancy is the lowest energy defect in these systems, and this defect is calculated to be shallow. Thus, in practice their Fermi energy is likely to lie close to their conduction band edges. The large electronegativity of the halogens means that the valence bands of these systems are very deep below the vacuum level. Even with $E_F$ lying at their conduction band edges, their work functions are still very large, much larger than that of the most electropositive metal, Pt.

We now pass to the case of $MoO_3$. This oxide has been widely used as a p-type dopant and electrode material in organic electronics, and has recently been used for p-type doping in carbon nanotubes, graphene and $MoS_2$ contacts. $MoO_3$ has two forms, the molecule $Mo_3O_9$, and a layered solid form MoO$_3$. MoO$_3$ was previously calculated to have a band gap of 3.0 eV and an electron affinity of 6.61 eV. Its oxygen vacancies were calculated to be shallow. The doping of MoS$_2$ and carbon nanotubes by MoO$_3$ layers has already been studied theoretically.

An orthorhombic supercell of graphene and MoO$_3$ was constructed as in Fig. 5(a). The electronic structure of the combined system was calculated. The large work function of MoO$_3$, 6.60 eV below that of graphene, means that there is a strong transfer doping. It is found that the Fermi energy of the combined system has shifted downwards in the graphene by 0.63 eV. In this case, doping has occurred by the transfer of electrons from the graphene valence band to the MoO$_3$ conduction band states. Nevertheless, the bonds between graphene and the outer O layer of MoO$_3$ are only physisorptive with a bond length of 2.5Å. The MoO$_3$ does not cause any puckering of the graphene sp$^2$ sites and thus does not affect the π bonding of the graphene layer. Thus, the C atoms do not acts as defects under this doping process. There will be no Raman D peak, and no carrier scattering. This is consistent with experiment where singularly Chen et al.[16] find that MoO$_3$ doped graphene retains the ability to show a quantum Hall effect, indicating a high carrier mobility.

MoO$_3$ is a very valuable dopant of graphene because it is a stable dopant, it raises the carrier concentration, it does not degrade the carrier mobility by causing defects, it has a wide band gap so that it is also optically transparent, a very useful combination useful for optical devices.

We now consider an n-type transfer dopant, CsO$_x$. Cs carbonate is widely used as an n-type dopant in organic light emitting diodes, and also can be used to dope graphene. The carbonate precursor dissociates on heating to leave a Cs oxide, which may actually be a sub-oxide. We consider the oxide to be Cs$_2$O. This has the inverse CdCl$_2$ hexagonal layered structure, with the Cs layers on the outside and O atoms on the inside. Note that whereas the interlayer bonding in CdCl$_2$ is van der Waals, the Cs-Cs bonding in Cs$_2$O is essentially metallic, not van der Waals. The hexagonal layers are reasonably lattice-matched to those of graphene, with a 1.6% mismatch, as shown in Table. 1 and Fig. 6(a). The Cs and O sites lie over the hollow sites of the graphene lattice.

Fig. 6(c) and (d) shows the band structure of isolated Cs$_2$O. Cs$_2$O is a semiconductor with a band gap of 1.44 eV in screened exchange and a very low electron affinity. Its valence band consists of oxygen 2p states. The valence band is very narrow because the O sites are far apart, so the O-O interface controlling the VB width is week.

Fig. 6(e) shows the density of states for the combined system. There is strong n-type doping, with electrons transferred from the Cs$_2$O valence band into the graphene conduction band. The E$_F$ of graphene is shifted upwards by 0.95 eV by the Cs$_2$O layer. Nevertheless the Cs-C bond is long and physisorbtive. It is not van der Waals, and no van der Waals correction to GGA is used in this case. The graphene atoms remain unpuckered below the Cs$_2$O and sp$^2$ bonding is maintained in the graphene. This behavior is consistent with the behavior of Cs2O as an n-type transfer dopant in organic semiconductors.[17,18]

Nitric acid is another p-type dopant, but it functions differently. Nistor et al.[32] studied the absorption of HNO$_3$ onto the graphene surface. They found that HNO$_3$ dissociates into an NO$_3$ radical, a NO$_2$ radical and a water molecule,

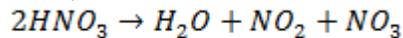
$$2HNO_3 \rightarrow H_2O + NO_2 + NO_3$$

HNO$_3$ is introduced into the 5x5 supercell. Dissociation occurs. These species are allowed to rotate to maximize their stability. The final geometry is shown in Fig. 7(a). The NO$_3$ radical lies planar parallel to the graphene plane, with each of its atoms lying on top of a carbon atom. The NO$_2$ radical and the water molecule lie in a plane normal to the graphene plane, with the central N atom of NO$_2$ and central O atom of H$_2$O nose down towards the graphene, as in Fig. 7(b). These species are physisorbed onto the graphene, and the bond lengths are given in Table. 2. None of the species cause any buckling of the underlying graphene layer. The binding energy (Table. 2) is relatively small.

Whereas H$_2$O is a closed shell system, both NO$_3$ and NO$_2$ are radicals each with a half-filled orbital. Critically, the work function of this orbital is greater than that of the graphene, the state lies deeper below the vacuum level than the Fermi energy E$_F$ of graphene. Thus, they give a singly empty state lying below E$_F$. This leads to an electron transfer from the graphene into the two NO$_x$ species, filling their states, and thus causing a hole doping of the graphene. As the bond length is long, there is only partial charge transfer. The charge transfer is calculated to be -0.3e on the NO$_3$ and -0.25e on the NO$_2$. This lowers E$_F$ of the graphene to -1.10 eV, as shown in Fig. 7(c). The retention of planar sp$^2$ bonding in the C sites under the NO$_3$ and NO$_2$ physisorbed species means that this does not constitute a defect, there is no Raman D peak and no carrier scattering. This is consistent with experiment. L'Arsie finds no change in the D peak intensity experimentally.[19]

We now consider Cl$_2$. Cl$_2$ is a closed shell molecule with a single Cl-Cl bond. It has a filled pσ state at -12 eV, two filled pπ states and two filled pπ* states, followed by an empty σ* state above its Fermi energy. The Cl$_2$ molecule is physisorbed onto graphene, but it does not produce doping because it has no empty states below E$_F$ of graphene (Fig. 8). There is no doping because the empty σ* state is high in energy despite the electronegativity of Cl.

Following Cl$_2$, we consider the O$_2$ molecule. This molecule is calculated to physisorb in a configuration across a C-C bond, as in Fig. 9(a). Now, the O$_2$ molecule is geometrically the same as the Cl$_2$ molecule, but its π* states would be half-filled in the spin unpolarized condition. This configuration is unstable to against symmetry breaking to open up a band gap. This occurs by an antiferromagnetic ordering of the σ* spins, with the up-spin states lying below E$_F$ and the down-spins lying above the gap. For the combined O$_2$ on graphene system, the gap is small enough that the empty spin-down σ* state lie below E$_F$ of isolated graphene, so there is a sizable charge transfer doping of the graphene by O$_2$, as shown in Fig. 9(b).

Finally, we consider the –OH radical. The O-H bond creates a deep-lying filled σ state, and a high-lying empty σ* state. The other broken O bond makes the unpaired electron of the radical. As O is very electronegative, this p state lies well below E$_F$ of isolated graphene. More interesting, this p state is able to form a strong C-O bond to a carbon atom underneath, puckering the C atom out of the plane, and converting it into a sp$^3$ configuration (Fig. 10). Thus, there is charge transfer from the graphene. However the overall effect on conductivity will be poor because the defect states will lower the mobility.

Overall, except for OH, there various dopants studied are physisorbed, without puckering the underlying graphene. This occurs because of the strong intra-layer rigidity of graphene, and its resistance to out-of-plane deformation needed to form the fourth extra bond to a chemisorbing species. We now show that this also applies to transfer dopants on hydrogen terminated diamond surfaces.

**Discussion**.

The electron affinity and ionization potentials of the various dopant species can be calculated using dopant supercells as described in the Methods section. The Fermi level shifts (FLS) are compared with the ionization potentials (IP) in Table. 1. We see that there is monotonic variation of the FLS with the IPs. The largest p-type shift occurs for SbF5 and FeCl3 has the largest shift of the more common dopants $FeCl_3$, $AuCl_3$, and $HNO_3$.

The $SbF_5$, $FeCl_3$ and $AuCl_3$ species have remarkably large ionization potentials, if the band gaps are added to the work functions.

Experimentally, $FeCl_3$ is found to give the largest $E_F$ shift of the common dopants $FeCl_3$, $AuCl_3$, $MoO_3$ and $HNO_3$.[33,34] Also, the absence of a Raman D peak at 1350 cm$^{-1}$ confirms that $FeCl_3$, $MoO_3$ and $HNO_3$ do not give rise to basal plane defects,[17,19,33] and thus should not increase carrier scattering. Our calculations have a similar aim as those of Hu and Gerber.[31] For $FeCl_3$, our calculations are for the spin-polarized state using GGA+U whereas Liu *et al.*[35] used spin unpolarized state. We used a three times smaller supercell than in Zhan *et al.*[33] by rotating the x,y axes. For $HNO_3$ doping we found that the acid dissociates, as in Nistor *et al.*[32]

**Conclusions**

We have calculated the conditions required for charge transfer doping of graphene (sometimes called non-covalent doping). We find that the Fermi level shift in eV is proportional to the electron affinity of the acceptor species or ionization potential of the donor species. We have treated a wider range of dopant species that other groups. The doping mechanism is similar to that occurring in transfer doping of organic semiconductors.

We thank EPSRC grant EP/P005152/1 and CSC for support.

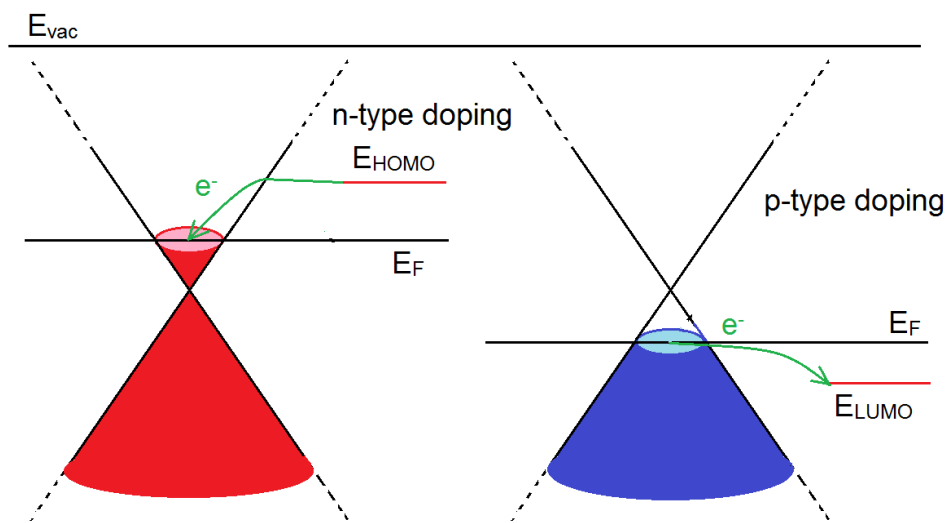

**Figure 1.** Schematic of n-type and p-type doping process in Graphene.

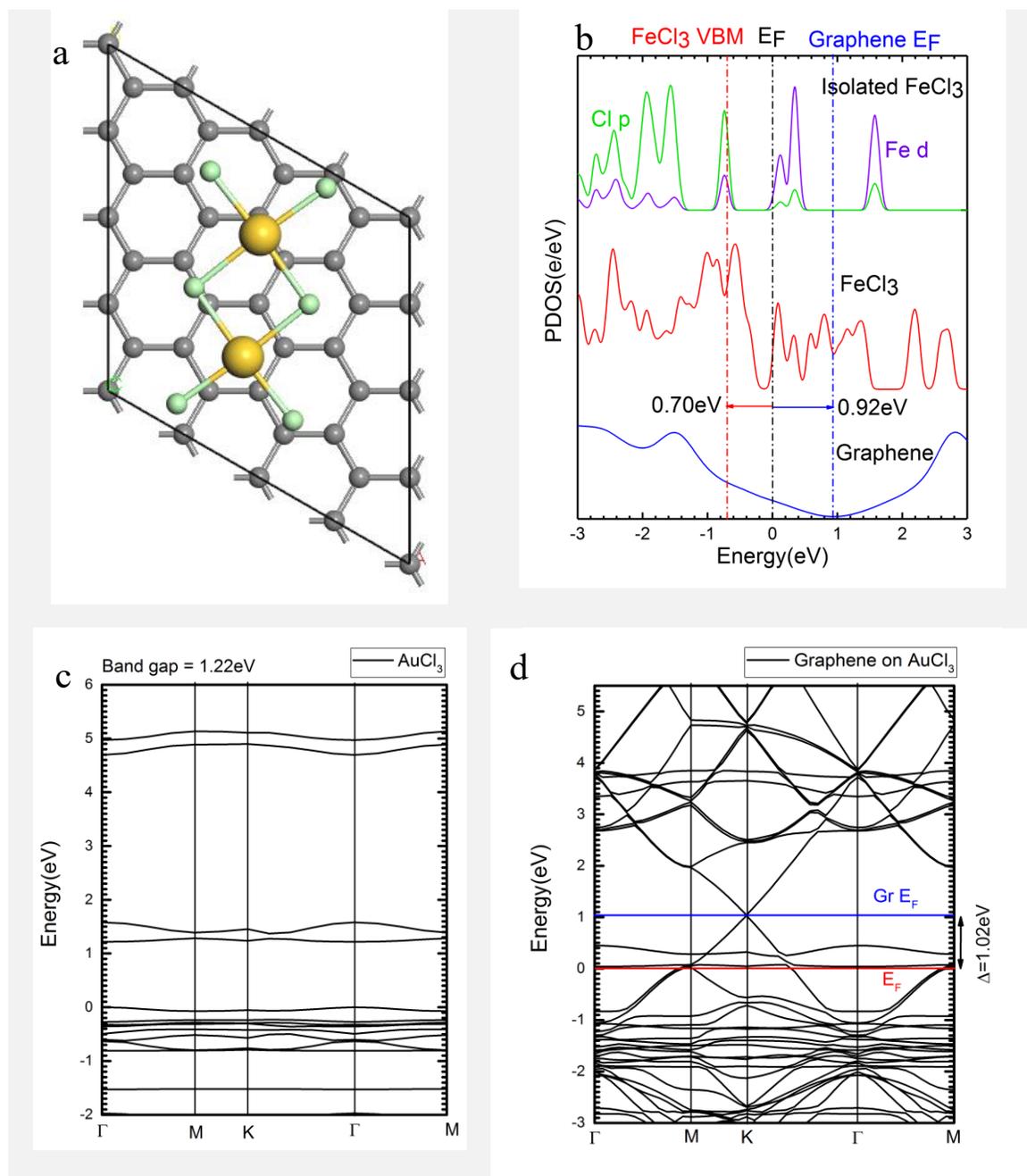

**Figure 2.** (a) Au$_2$Cl$_6$ molecular on 4x4 Graphene supercell. (b) PDOS of isolated AuCl$_3$ and AuCl$_3$/Graphene system. (c) Band Structure of isolated pure Au$_2$Cl$_6$ in the hexagonal lattice. (d) Band Structure of the combined system.

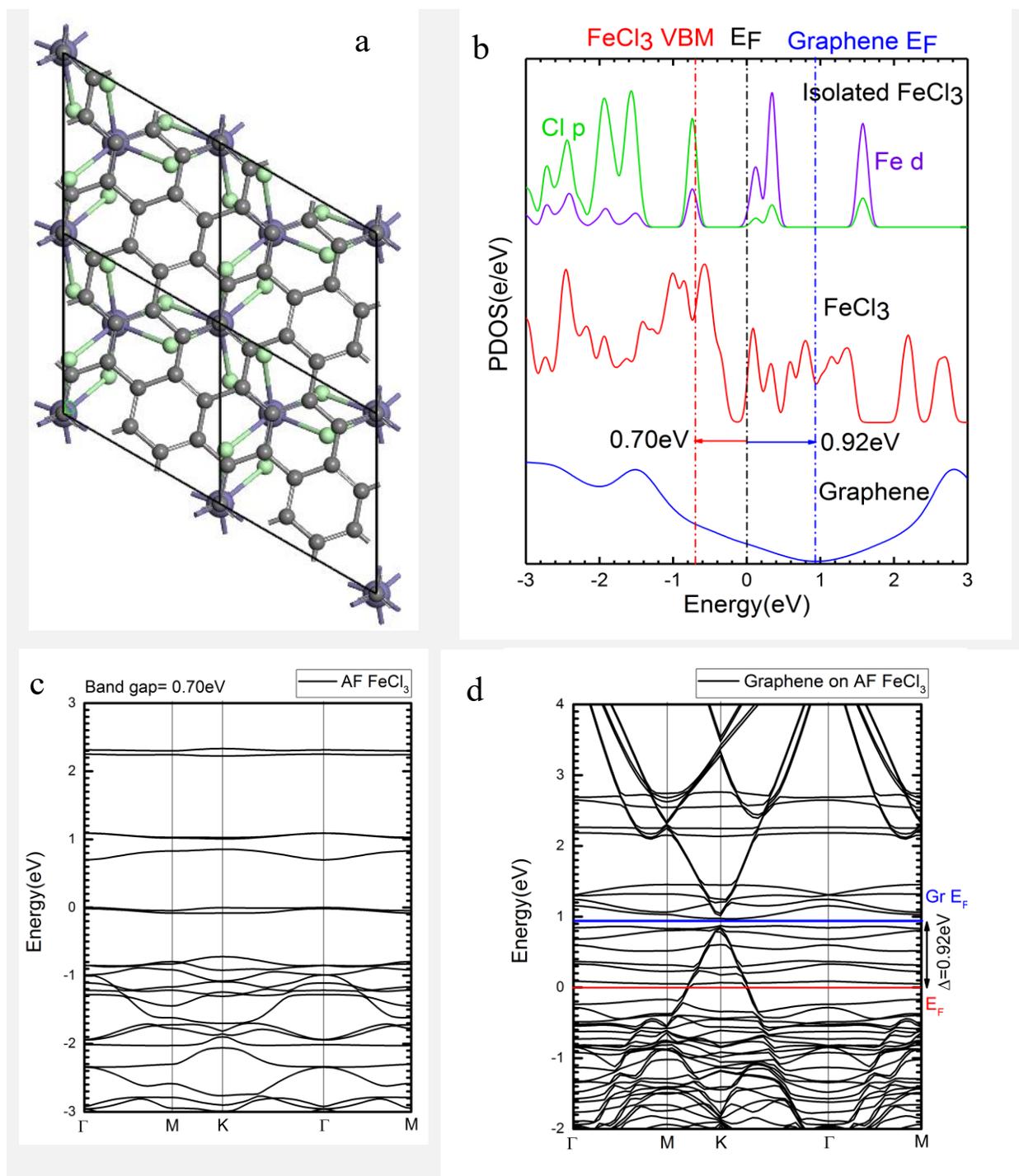

**Figure 3.** (a) FeCl$_3$ on √7x√7 Graphene supercell. (b) PDOS of isolated AF FeCl$_3$ and FeCl$_3$/Graphene system. (c) Band Structure of isolated pure AF FeCl$_3$. (d) Band Structure of the combined system.

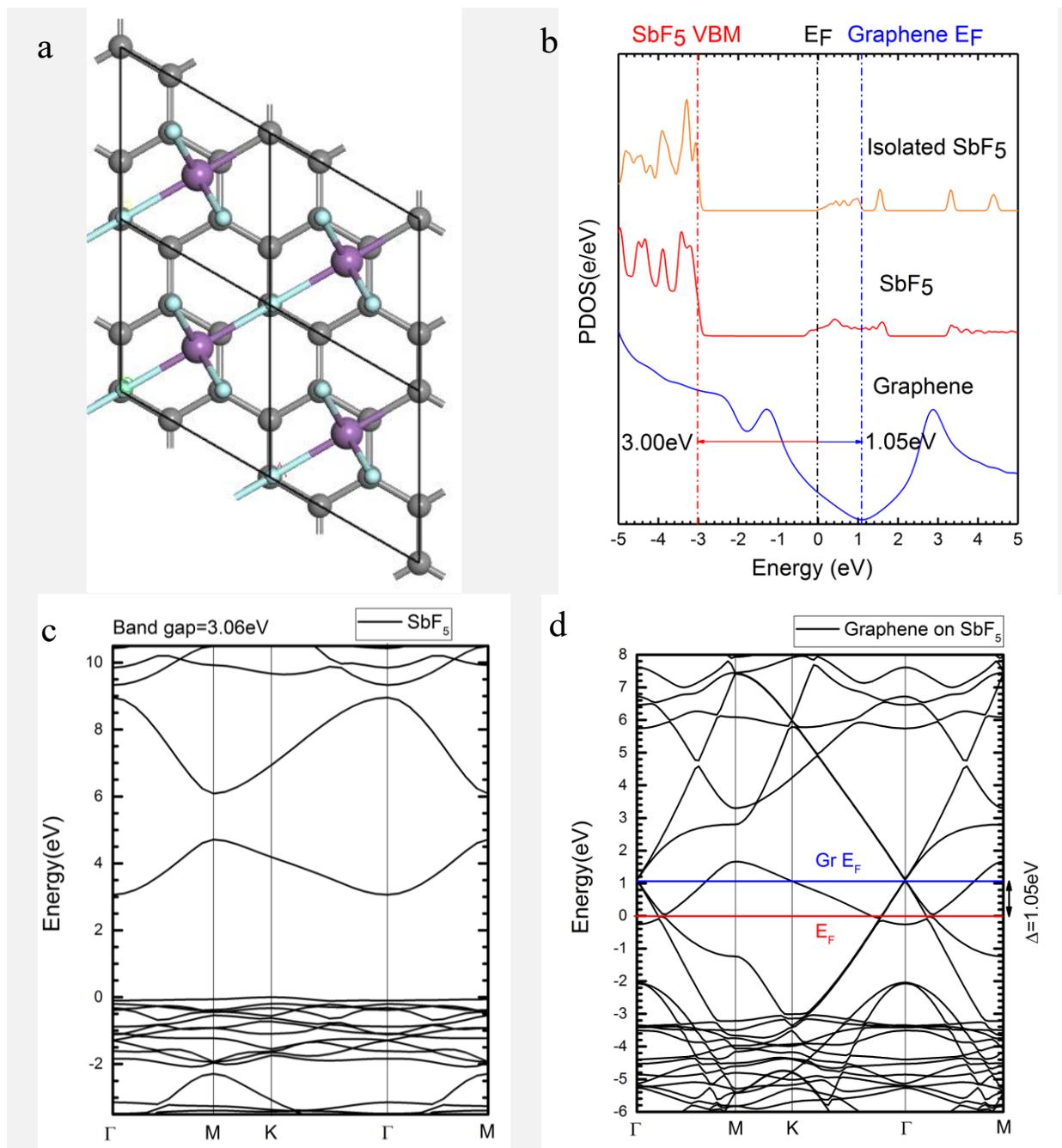

**Figure 4.** (a) SbF$_5$ on √3x√3 Graphene supercell. (b) PDOS of isolated SbF$_5$ and SbF$_5$/Graphene system. (c) Band Structure of isolated pure SbF$_5$ single layer in the hexagonal lattice. (d) Band Structure of the combined system.

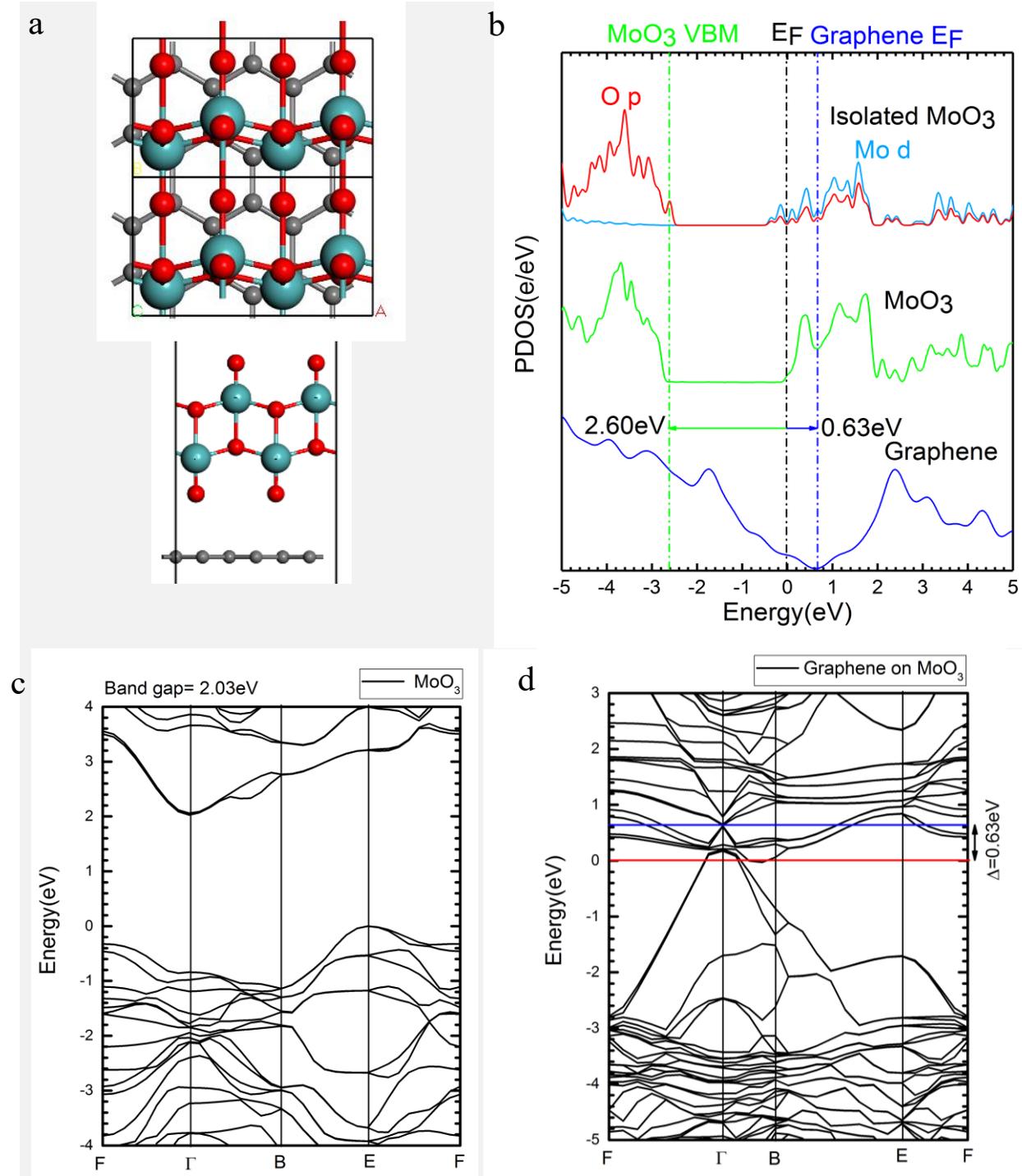

**Figure 5.** (a) MoO$_3$ on 3x√3 Graphene supercell. (b) PDOS of isolated MoO$_3$ and MoO$_3$/Graphene system. (c) Band Structure of isolated pure single layer MoO$_3$. (d) Band Structure of the combined system.

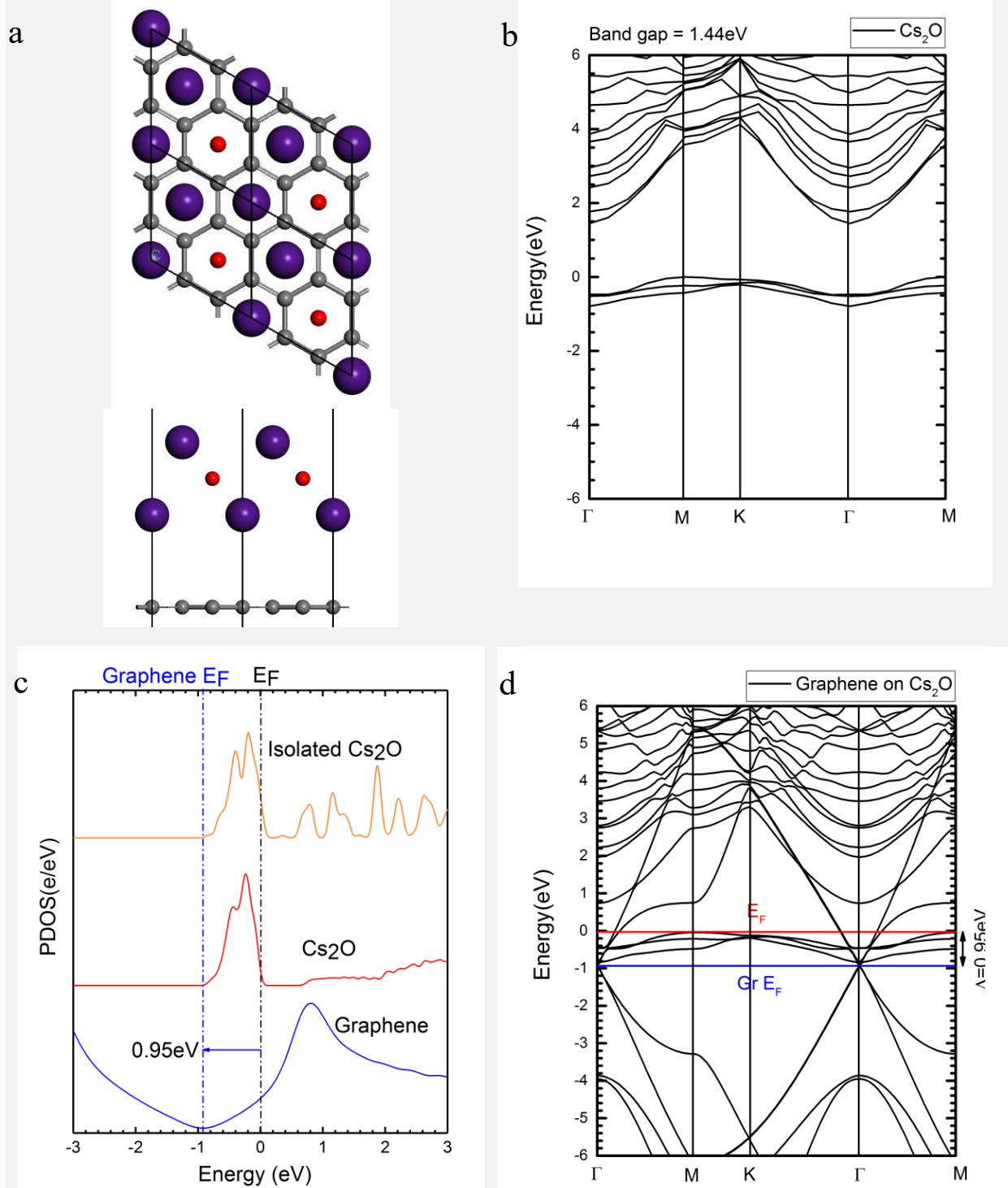

**Figure 6.** (a) Top view and (b) side view of $Cs_2O$ on $\sqrt{3} \times \sqrt{3}$ Graphene supercell. (c) GGA and (d) screened exchange Band Structure of isolated pure single layer $Cs_2O$. (e) PDOS of isolated $Cs_2O$ and $Cs_2O$/Graphene system. (f) Band Structure of the combined system.

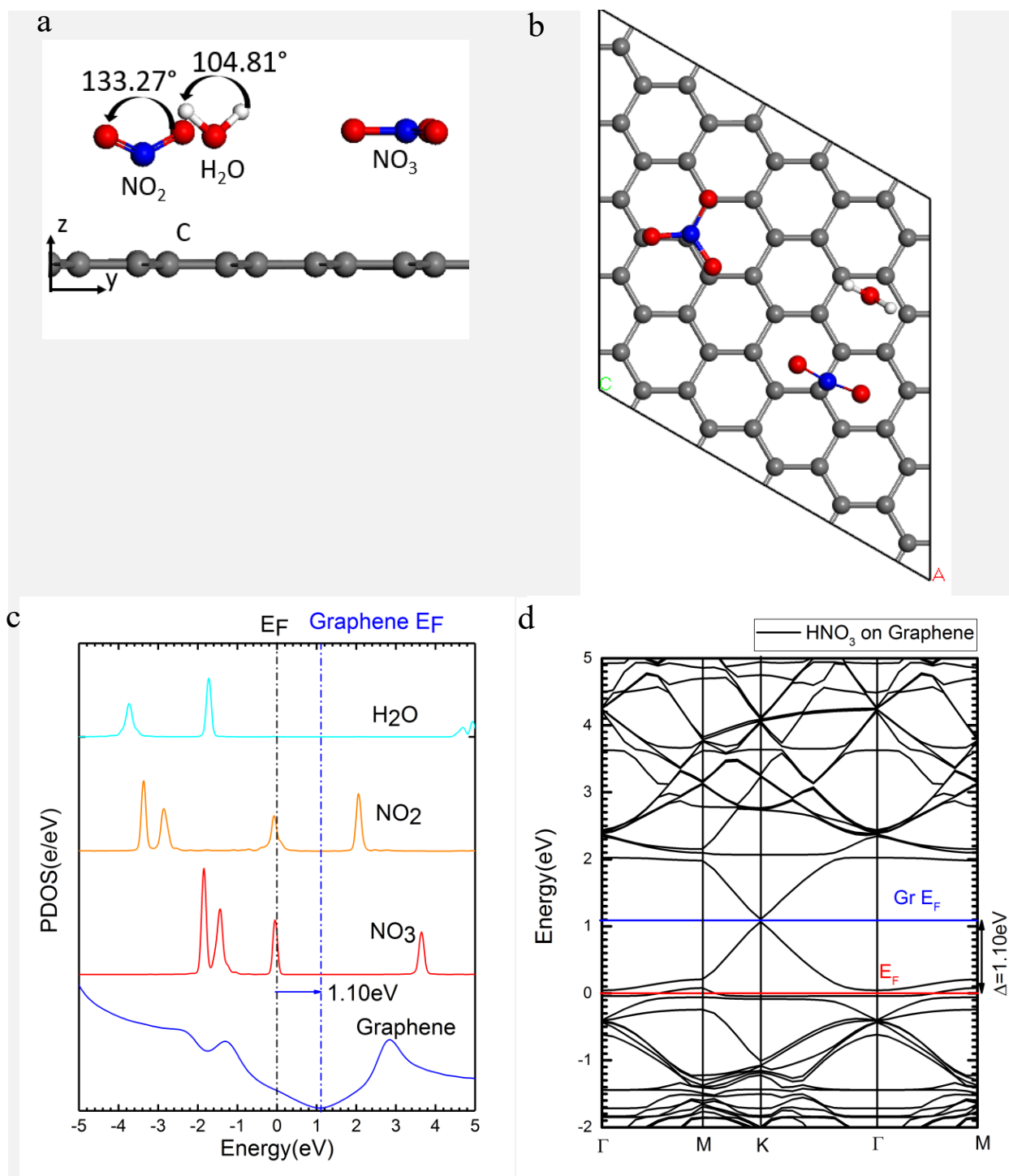

**Figure 7.** (a) Side view and (b) top view of 2HNO$_3$ dissolve onto 5x5 Graphene supercell. (c) PDOS of 2HNO$_3$/Graphene system. (d) Band Structure of the combined system.

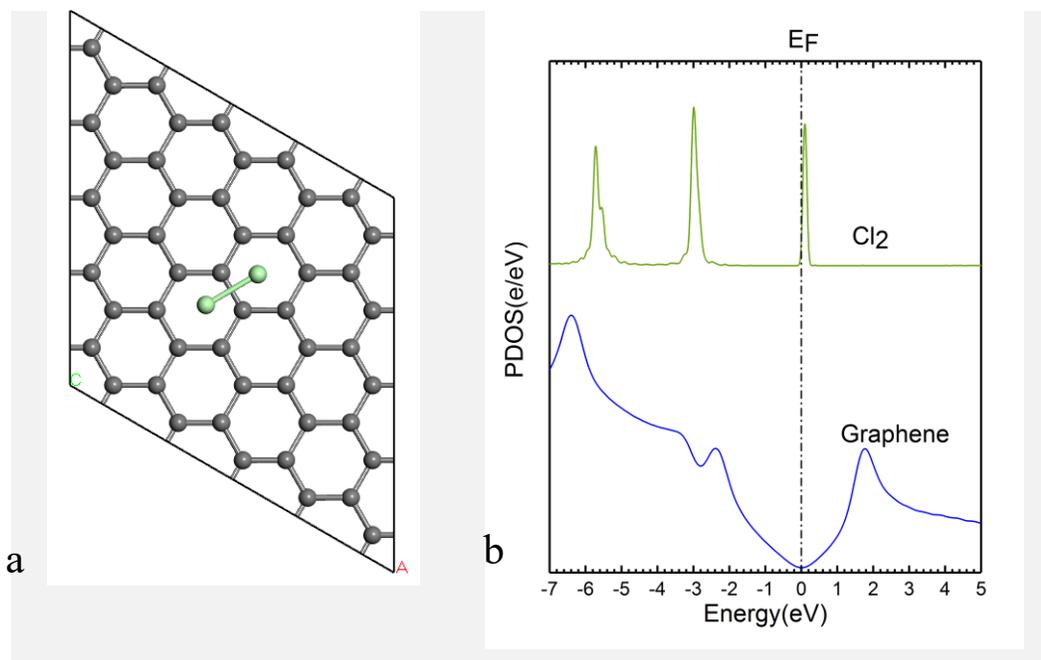

**Figure 8.** (a) Top view of Cl$_2$ on 5x5 Graphene supercell. (b) PDOS of Cl$_2$/Graphene system.

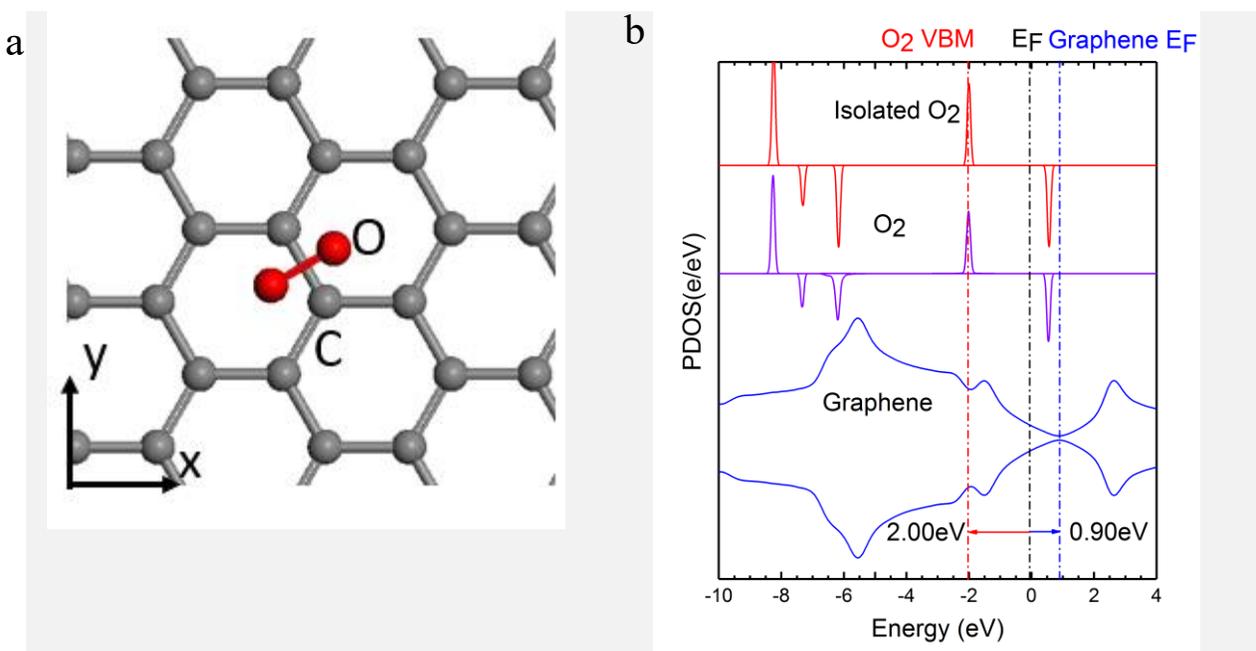

**Figure 9.** (a) Top view of triplet O$_2$ on 5x5 Graphene supercell. (b) PDOS of O$_2$/Graphene system.

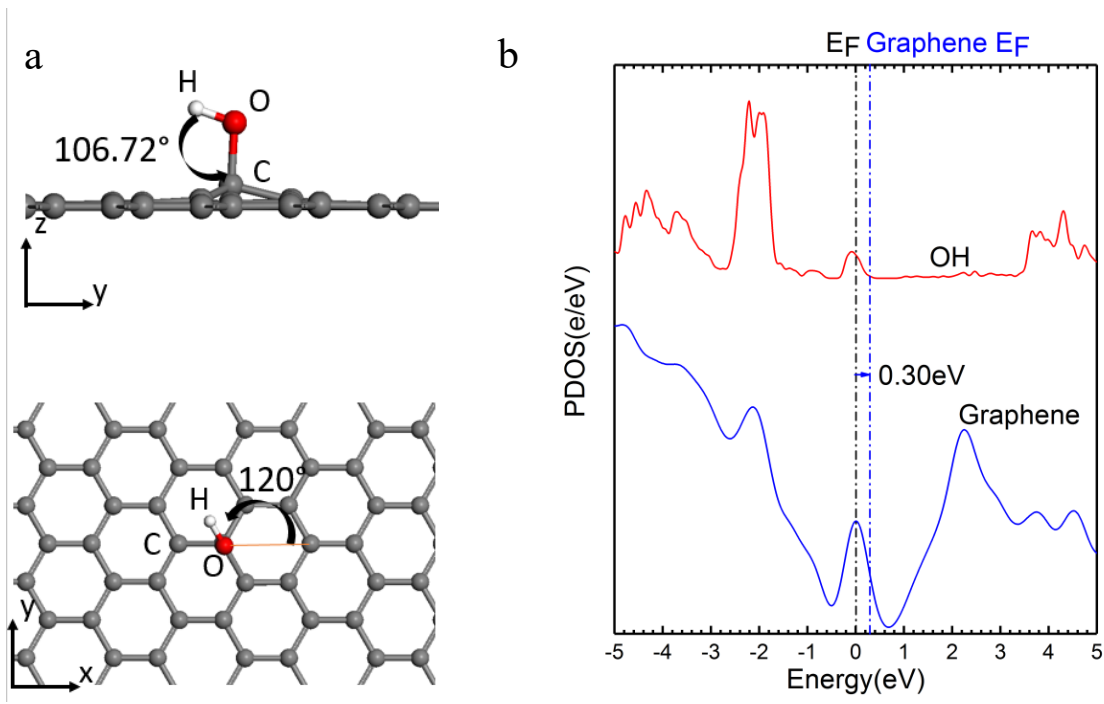

**Figure 10.** (a)Side view and (b) top view of OH radical bonding onto 5x5 Graphene supercell. (c) PDOS of OH/Graphene system.

**Table 1.** Supercell and lattice match of Graphene and dopant. M in the mismatch column refers to a molecular dopant where there is no mismatch.

|  | Graphene supercell/ Dopant supercell | Mismatch rate (%) |
|---|---|---|
| SbF$_5$ | $\sqrt{3} \times \sqrt{3}/1 \times 1$ | 1.66 |
| FeCl$_3$ | $\sqrt{7} \times \sqrt{7}/1 \times 1$ | 1.42 |
| AuCl$_3$ | $4 \times 4/1 \times 1$ | M |
| MoO$_3$ | $3 \times \sqrt{3}/2 \times 1$ | 0.34, 7.92 |
| Cs$_2$O | $\sqrt{3} \times \sqrt{3}/1 \times 1$ | 1.62 |
| Cl$_2$ | $5 \times 5$ | M |
| O$_2$ | $5 \times 5$ | M |
| OH | $5 \times 5$ | M |
| HNO$_3$ | $5 \times 5$ | M |

**Table 2.** Atomic distance, bond length and puckering of Graphene.

|  | Bond type | Bond (Å) | Surface distance (Å) | Puckering (Å) | Binding energy (eV) |
|---|---|---|---|---|---|
| OH | O-H | 0.98 | — | 0.51 | -1.64 |
|  | C-O | 1.51 | — |  |  |
| O$_2$ | O-O(in O$_2$) | 1.24 | 3.29 | 0.09 | -0.13 |
| HNO$_3$ | O-H(in H$_2$O) | 0.98 | 3.28 | 0.06 | -0.39 |
|  | N-O(in NO$_2$) | 1.23 | 2.60 |  |  |
|  | N-O(in NO$_3$) | 1.27 | 3.25 |  |  |

**Table 3.** Atomic distance, bond length and puckering of Graphene. Layer distance, work function, ionization potential and Fermi level shift.

|  | Layer distance (Å) | Work Function (eV) | Ionization potential (eV) | FLS (eV) |
|---|---|---|---|---|
| **SbF$_5$** | 3.65 | 7.04 | - | -1.05 |
| **FeCl$_3$** | 3.54 | 6.97 | - | -0.92 |
| **MoO$_3$** | 2.95 | 6.61 | - | -0.63 |
| **AuCl$_3$** | 3.35 | 5.94 | - | -1.02 |
| **Cs$_2$O** | 3.75 | - | 2.35 | 0.95 |